\documentclass[conference]{IEEEtran}
\IEEEoverridecommandlockouts
\usepackage{cite}
\usepackage{amsmath,amssymb,amsfonts}
\usepackage{algorithmic}
\usepackage{graphicx}
\usepackage{textcomp}
\usepackage{xcolor}
\usepackage{tikz}
\usepackage{graphicx}
\usepackage{amssymb}
\usepackage{amsmath,amssymb,amsfonts,amsthm}
\usepackage{mathtools}
\usepackage{graphicx}
\usepackage{hyperref}
\usepackage{tikz}

\def\BibTeX{{\rm B\kern-.05em{\sc i\kern-.025em b}\kern-.08em
    T\kern-.1667em\lower.7ex\hbox{E}\kern-.125emX}}
\begin{document}

\title{A Python Tool for Object-Centric Process Mining Comparison (Extended Abstract)\\}

\author{\IEEEauthorblockN{Anahita Farhang Ghahfarokhi}
\IEEEauthorblockA{\textit{Process and Data Science (Informatik 9)} \\
\textit{RWTH Aachen University}\\
 Aachen, Germany \\
farhang@pads.rwth-aachen.de}
\and
\IEEEauthorblockN{Wil M.P. van der Aalst}
\IEEEauthorblockA{\textit{Process and Data Science (Informatik 9)} \\
\textit{RWTH Aachen University}\\
 Aachen, Germany \\
wvdaalst@rwth-aachen.de}
}

\maketitle

\begin{abstract}
Object-centric process mining provides a more holistic view of processes where we analyze processes with multiple case notions. However, most object-centric process mining techniques consider the whole event log rather than the comparison of existing behaviors in the log. In this paper, we introduce a stand-alone object-centric process cube tool built on the PM4PY-MDL process mining framework. Our infrastructure uses both object and event attributes to build the process cube which leads to different types of materialization. Furthermore, our tool is equipped with the state of the art object-centric process mining techniques. Through our tool the user can visualize the extracted object-centric event log from process cube operations, export the object-centric event log, discover the state-of-the-art object-centric process model for the extracted log, and compare the process models side-by-side.
\end{abstract}

\begin{IEEEkeywords}
Object-Centric Process Mining, Object-Centric Event Logs, Process Comparison, Process Cubes
\end{IEEEkeywords}
\vspace{-0.09cm}
\section{Introduction}
Process mining is a field of data science that aims to bridge the gap between business process model-based analysis and data-oriented analysis. Process mining techniques include process discovery, conformance checking, and process enhancement methods~\cite{uysal2020process}.

Event logs are the starting point to apply process mining techniques. Event logs consist of events where each event refers to one case notion, activity, timestamp, and some additional attributes such as resource. Common process mining techniques are based on event logs with one case notion, however, in reality, several case notions are involved in one process, e.g., a simple Order-to-Cash (O2C) process where orders, offers, and invoices are involved. Object-centric process mining is a novel branch of process mining that aims to develop process mining techniques on top of Object-Centric Event Logs (OCELs)~\cite{ghahfarokhi2021ocel}. Initial approaches have been developed to extract OCEL logs from information systems~\cite{simovic2018domain,berti2018extracting,gonzalez2019process} and discover process models from OCEL logs~\cite{cohn2009business,nooijen2012automatic}. However, there may exist a variety in object-centric processes that requires the separation of different processes from each other~\cite{ghahfarokhi2021process}. Therefore, process cubes are introduced that are inspired by the notion of OLAP and are developed to compare processes with each other through process cube operations such as slice and dice~\cite{vogelgesang2013multidimensional}. Several implementations of process cubes are developed ~\cite{ribeiro2011event, bolt2015multidimensional}. However, they cannot support event logs with multiple objects, i.e., OCEL logs. 

In this demo paper, we present an interactive tool that permits the user to compare object-centric processes with each other through process cube operations. Furthermore, it permits to discover object-centric process models (object-centric Directly Follows Graphs~\cite{berti2018extracting}, Object-Centric Petri Nets~\cite{van2020discovering}) and compare the process models side-by-side.

The remainder of this paper is organized as follows. In Section~\ref{Object-Centric Process Cube}, we describe the main functionalities of the tool that are provided. In Section~\ref{scalability}, we evaluate the scalability of the tool. Finally, Section~\ref{conclusion} concludes the paper
and presents some future work for the extension of our work.
\begin{figure*}[!]
    \vspace{-1cm}
	\centering
		\includegraphics[scale=.4]{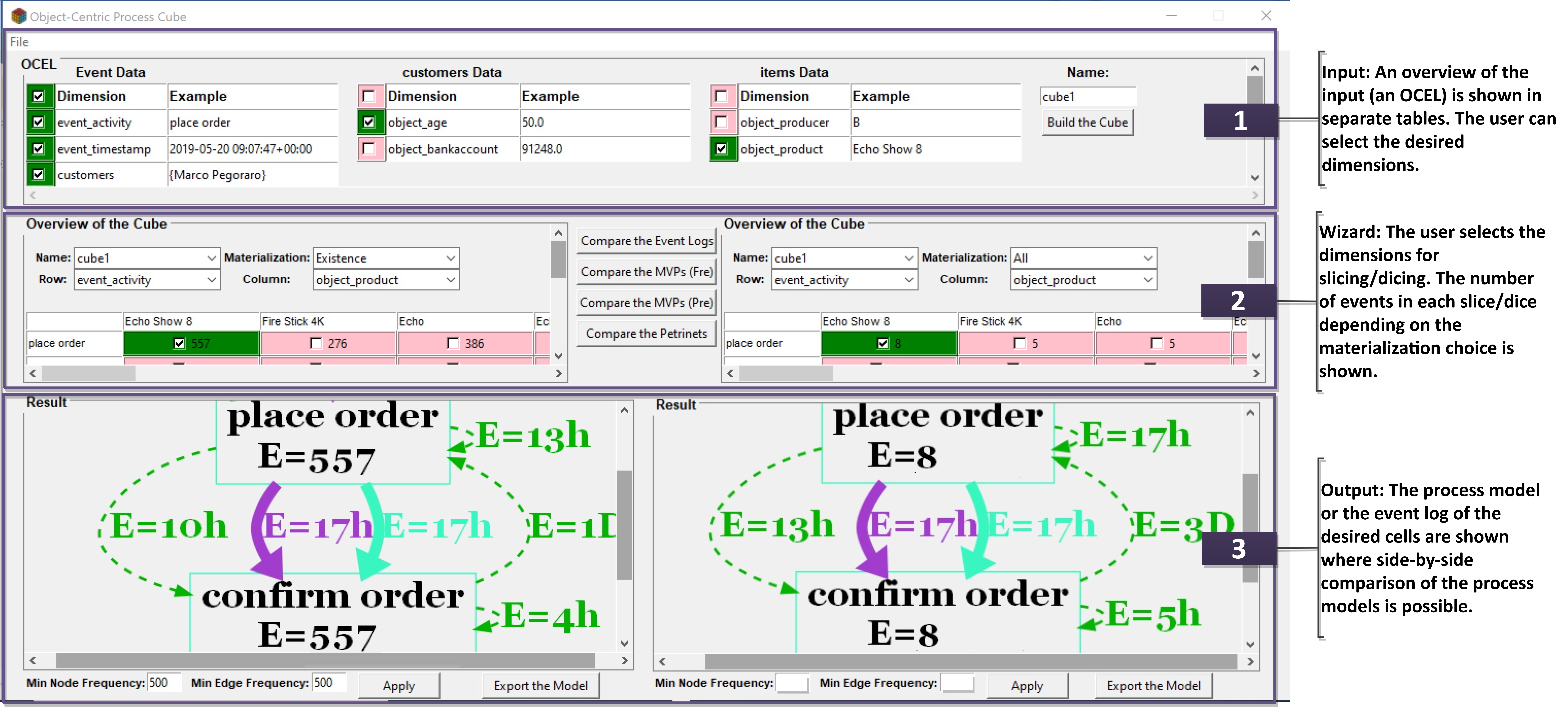}
	\caption{The user interface of the tool that is developed in Tkinter GUI.}
	\label{UItool}
\end{figure*}

\section{Object-Centric Process Cubes}\label{Object-Centric Process Cube}
In our open-source tool\footnote{\url{https://github.com/AnahitaFarhang/object-centric-process-cube}}, called OCPC (Object-Centric Process Cube), we have used Tk GUI toolkit as the user interface which is one of the popular standard Python interfaces. A snapshot of the tool is shown in Figure~\ref{UItool} where three sub~windows are highlighted:
\begin{itemize}
 \item Input: the input is JSON-OCEL/XML-OCEL. In OCEL, an event is related to event attributes and some possible objects related to that event. Furthermore, each object can have its own properties in another table. Therefore, as shown in Figure~\ref{UItool}, in the first highlighted sub~window, we have shown events with event attributes and their objects in one table, whereas, objects within their attributes are shown in separate tables. The user can select event attributes and object attributes as the dimensions of the process cube and build the cube. 
 \item Wizard: After creating the cube, we see overviews of the cube in the second sub~window. It is possible to select dimensions of the cube as rows and columns and see the number of events in the selected slices/dices.
 
 The combo box "Materialization" is related to design choices that we have in this tool for materializing that did not exist in the previous process cubes. In the "Existence" option for materialization, there should exist an object in that event that satisfies the property. For example, in the left-view in Figure~\ref{UItool},  the option is "Existence" and the highlighted green dice in this view shows in 557 events \emph{Echo~show~8} was involved. In the "All" option for materialization, all the objects, in the event, should satisfy that property. An example is shown in right-view where this option is "All". The highlighted green cell in this view shows that in 8 events all the items were \emph{Echo~show~8}. 
 \item Output: In the third sub~window, we have compared the process models of the selected cells shown in the second sub~window. As we see, there are differences in the duration of activities for the selected cells. Furthermore, depending on the operation selection in the second sub~window, it is possible to compare the extracted event logs, and Object-Centric Petri nets of the selected cells.
\end{itemize}

\vspace{-0.1cm}
\section{Scalability}\label{scalability}

The scalability of the tool in terms of number of events, event attributes, and object attributes are shown in Tables~\ref{nevents},~\ref{neventatts}, and~\ref{nobjectatts}, respectively\footnote{All the above experiments were performed on a laptop with the specifications: PC Intel(R) Core(TM) i7-8550U CPU @ 1.80GHz.}. The result of the analysis with different settings shows the time required for creating the cube increases linearly, linearly, and non-linearly when increasing the number of events, event attributes, and object attributes, respectively. These relationships are justifiable by the nature of OCELs where an event can contain multiple objects and there is a one-to-many relationship between an event and its objects which results in a non-linear relationship shown in Table~\ref{nobjectatts}.

\begin{table}[!]
\vspace{-0.2cm}
\caption{Scalability of the tool for different numbers of events: n\_event attributes{=}4, and n\_object attributes{=}4}
\centering
\scalebox{0.85}{%
\begin{tabular}{|l|l|l|l|l|l|}
\hline
Number of events          & 1000  & 5000   & 10000  & 15000  & 20000  \\ \hline
Time of building the cube (s) & 11.15 & 120.23 & 237.26 & 361.44 & 490.12 \\ \hline
\end{tabular}}
\label{nevents}
\vspace{-0.3cm}
\end{table}

\begin{table}[!]
\caption{Scalability of the tool for different numbers of event attributes: n\_events{=}20000, and n\_object attributes{=}4}
\centering
\scalebox{0.85}{%
\begin{tabular}{|l|l|l|l|l|}
\hline
Number of object attributes          & 0  & 1   & 2  & 3  \\ \hline
Time of building the cube (s) & 60.17 & 127.27 & 254.79  & 490.12 \\ \hline
\end{tabular}}
\label{neventatts}
\vspace{-0.1cm}
\end{table}

\begin{table}[!]
\vspace{-0.2cm}
\caption{Scalability of the tool for different numbers of object attributes: n\_events{=}20000, and n\_event attributes{=}4}
\centering
\scalebox{0.85}{%
\begin{tabular}{|l|l|l|l|l|l|}
\hline
Number of object attributes          & 0  & 1   & 2  & 3  & 4 \\ \hline
Time of building the cube (s) & 62.16 & 99.43 & 233.63 & 270.64 & 490.12 \\ \hline
\end{tabular}}
\label{nobjectatts}
\vspace{-0.4cm}
\end{table}

\section{Conclusion}\label{conclusion}
Here, we present an interactive object-centric process cube tool that enables the exploration, discovery of
object-centric process models along with side-by-side model comparison. Using the tool
that is implemented in PM4PY-MDL, users explore object-centric event logs using
process cube operations, export the partitioned OCEL, and discover the object-centric process models, annotated with frequency and performance, which helps in understanding processes better. A video displaying the functionalities of our tool is available at the address \url{https://youtu.be/zenHt3wdZP4}. We aim in the future to extend this tool with an automatic method to choose interesting slices and dices to explore the cube.

\bibliographystyle{plain}
\bibliography{bibliography}

\end{document}